\documentclass[12pt,a4paper]{article}
\usepackage[margin=1.02in]{geometry}
\usepackage{amsmath,amssymb,amsthm,mathtools}
\usepackage{hyperref}
\usepackage{xcolor}
\usepackage{cite}
\usepackage{microtype}
\usepackage{graphicx}

\hypersetup{colorlinks=true,linkcolor=blue!55!black,citecolor=blue!55!black,urlcolor=blue!55!black}

\newcommand{\dd}{\mathrm{d}}
\newcommand{\ii}{\mathrm{i}}
\newcommand{\ee}{\mathrm{e}}

\newcommand{\IR}{\mathrm{IR}}
\newcommand{\dU}{d_{\mathcal U}}
\newcommand{\dEff}{d_{\mathrm{eff}}}

\newcommand{\calP}{\mathcal P}
\newcommand{\calF}{\mathcal F}
\newcommand{\rhoinc}{\rho_{\mathrm{inc}}}

\theoremstyle{plain}
\newtheorem{theorem}{Theorem}
\newtheorem{proposition}{Proposition}

\theoremstyle{remark}

\title{\textbf{The Infraparticle Edge}}
\author{\sc Soo-Jong Rey\\[0.2cm]
{\sl Kwangwoon University}\\
{\sl Seoul, Korea}}
\date{\tt sjrey@kw.ac.kr}

\begin{document}
\maketitle

\begin{abstract}
I derive the charged-particle spectral edge from the quantum instrument of soft QED. I use two projections of that instrument.  Tracing over unresolved photons gives the reduced hard-sector channel.  Pushing the outcomes to total energy gives the inclusive energy distribution.  Its Laplace exponent is fixed by the diagonal soft intensity.  For
${\rm d} N_h(\omega)=\eta_h{\rm d}\omega/\omega+{\rm d} N_{h,\mathrm{reg}}(\omega)$,
I obtain
$\rho_{\mathrm{inc}}(s)\sim C\theta(s-m^2)(s-m^2)^{-1+\eta_h}$.
I retain the coherence kernel and derive hard-sector dephasing and the spectral edge from two contractions of one soft environment.  The diagonal coefficient $\kappa_{aa}$ fixes the endpoint exponent, while
$\frac12(\kappa_{aa}+\kappa_{bb}-2\operatorname{Re}\kappa_{ba})$
fixes the dephasing exponent between hard alternatives.  I then classify infrared energy marginals, derive the finite-resolution residue
$Z(\mu)=(\mu/\Lambda)^{\eta_h}$, prove stability under infrared-integrable perturbations, and separate the bath exponent from a hard threshold exponent.  For the one-electron spectral measure, the hard threshold factor is regular.  The resulting edge has the local power law of a gapped unparticle spectrum, while its exponent remains a response coefficient of the unresolved photon sector.
\end{abstract}

\section{Introduction}

A stable massive particle contributes an atom to the Kallen--Lehmann spectral measure \cite{Kallen1952,Lehmann1954}:
\begin{equation}
  \rho(s)=Z\delta(s-m^2)+\rho_{\rm cont}(s),\qquad Z>0.
  \label{eq:particle-atom}
\end{equation}
In QED, arbitrarily soft photons~\cite{Weinberg1965} accumulate at the charged threshold.  The threshold remains at \(s=m^2\), while the atomic weight flows into a continuum.  The exact charged sector carries an infraparticle edge rather than an isolated mass shell \cite{BlochNordsieck1937,Buchholz1986,Buchholz1987}.
In this paper, I derive this edge from the open-quantum-system formulation of soft QED I developed in a series of papers \cite{PaperA,PaperB,PaperC}.  I take hard charged data as the system and unresolved photons as the environment.  The resulting quantum instrument assigns an outcome operator to each soft-photon configuration.  I use that instrument in two ways.  First, I trace over all outcomes and recover the reduced hard-sector channel.  Second, I group the outcomes by total soft energy and obtain the probability law that controls the spectral endpoint.

Let \(X\) denote an unresolved photon configuration, let \(E(X)\) be its total energy below a separation scale \(\Lambda\), and let \(h\) label the hard charged configuration.  The energy marginal has Laplace transform
\begin{equation}
  \widetilde{\calP}_h(\tau)
  =
  \int \dd{\mathbb P}_h(X)\,\ee^{-\tau E(X)}
  =
  \exp\left[
  \int_0^\Lambda \dd N_h(\omega)\bigl(\ee^{-\tau\omega}-1\bigr)
  \right].
  \label{eq:paperA-input}
\end{equation}
The measure \(\dd N_h\) is the diagonal soft intensity.  QED gives
\begin{equation}
  \dd N_h(\omega)=\eta_h\frac{\dd\omega}{\omega}+\dd N_{h,{\rm reg}}(\omega),
  \qquad
  \int_0^\Lambda \dd N_{h,{\rm reg}}(\omega)<\infty .
  \label{eq:qed-intensity}
\end{equation}
The coefficient \(\eta_h\) fixes the endpoint power.  The integrable part fixes the normalization and regular terms.  The infraparticle branch cut is established infrared physics.  I use it here to expose the spectral information already contained in the open-system instrument and to classify the resulting endpoint.

The reduced channel probes the off-diagonal soft data.  Two hard alternatives \(a\) and \(b\) emit conditional soft amplitudes \(s_a\) and \(s_b\).  Their dephasing rate depends on \(|s_a-s_b|^2\), while the spectral edge for \(a\) depends on \(|s_a|^2\).  I keep the two exponents separate.  They share one environment and represent different observables.

I import the quantum instrument and soft kernel from previous works~\cite{PaperA,PaperB,PaperC}.  I then change the environmental readout from ``which hard alternative'' to ``how much unresolved energy.''  The first readout gives decoherence; the second gives the endpoint distribution.  The calculation starts from an open-quantum-system map and ends with a Kallen--Lehmann measure.

I compare the result with the unparticle spectral law \cite{Georgi2007,Georgi2007Another}:
\begin{equation}
  \rho_{\mathcal U}(P^2)=A_{\dU}\,\theta(P^0)\theta(P^2)(P^2)^{\dU-2}.
  \label{eq:georgi-rho}
\end{equation}
The QED charged sector instead has a massive endpoint,
\begin{equation}
  \rhoinc(s)\sim C\,\theta(s-m^2)(s-m^2)^{-1+\eta_h}.
  \label{eq:qed-edge-intro}
\end{equation}
Both laws contain a continuum and a noninteger power.  The unparticle power follows from operator scaling; the QED power follows from the unresolved photon measure attached to a massive threshold.

\begin{figure}[t]
  \centering
  \includegraphics[width=0.95\linewidth]{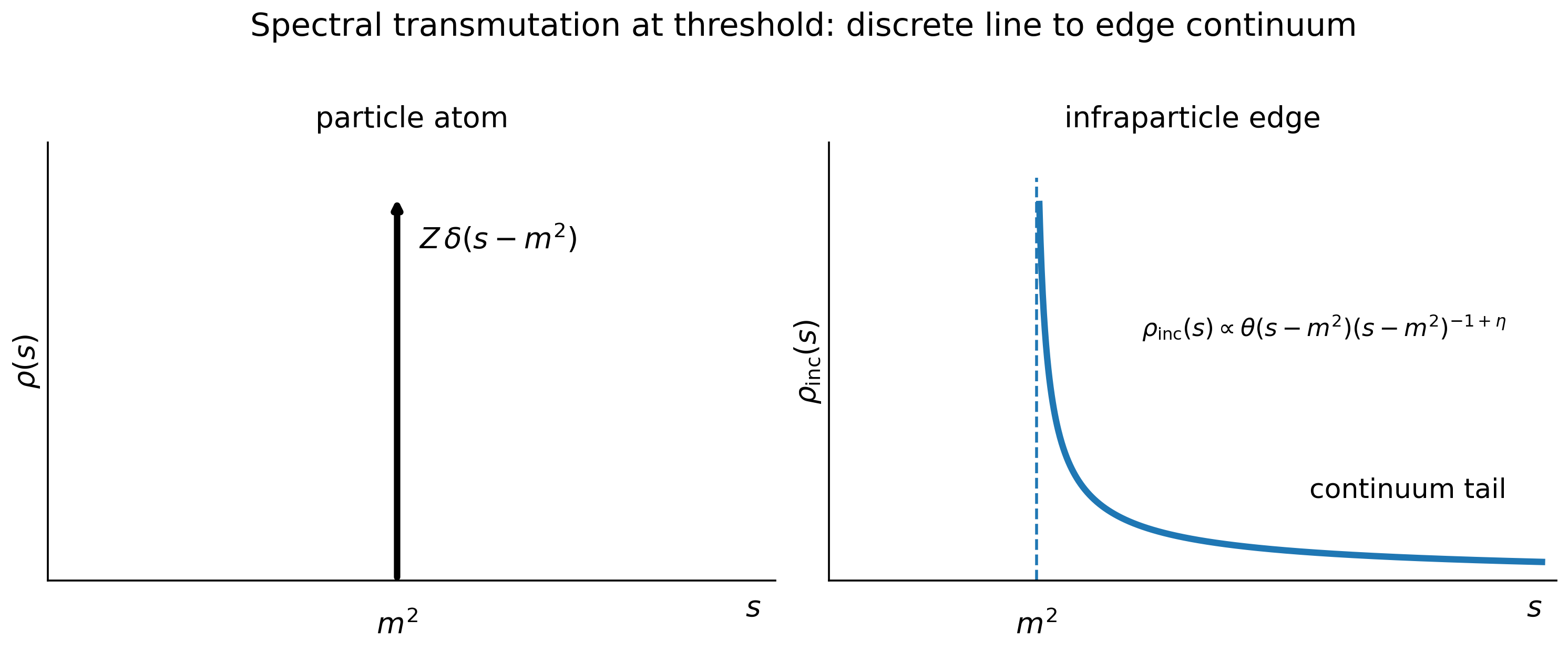}
  \caption{A stable particle contributes an isolated atom at \(s=m^2\).  The charged QED sector begins at the same threshold and carries an integrable continuum edge.  The unresolved photon environment transfers the atomic weight into the continuum.}
  \label{fig:edge-schematic}
\end{figure}

I plan the rest of paper as follows. I proceed from the soft-QED instrument to its coherence kernel, energy marginal, and spectral edge.  I then separate the soft and hard threshold exponents, compare the result with unparticle spectra, and derive the infrared classification and residue flow.  The appendices collect the asymptotic calculations.

\section{Soft QED as an open quantum system}

Fix a hard charged configuration and a soft scale \(\Lambda\).  An unresolved outcome is a photon configuration
\begin{equation}
  X=\{(k_1,\lambda_1),\ldots,(k_n,\lambda_n)\},
  \qquad |k_i|<\Lambda .
  \label{eq:soft-outcome-X}
\end{equation}
I denote its instrument element by \(\mathsf I_h[\dd X]\).  Inclusive normalization gives
\begin{equation}
  \int \mathsf I_h[\dd X]=1.
  \label{eq:instrument-normalization-review}
\end{equation}
This normalization combines the zero-photon contribution with all real unresolved sectors.  More explicitly, the instrument acts on a hard density matrix as
\begin{equation}
  \mathcal I[\dd X](\rho)=K_X\rho K_X^\dagger,
  \qquad
  p_\rho(\dd X)=\operatorname{Tr}\mathcal I[\dd X](\rho).
  \label{eq:instrument-map}
\end{equation}
The map is completely positive for every outcome set, and the integral over all outcomes is trace preserving.

Tracing over the soft outcome gives the hard-sector channel
\begin{equation}
  \rho_{\rm hard}\longmapsto
  \Phi(\rho_{\rm hard})
  =\int K_X\rho_{\rm hard}K_X^\dagger .
  \label{eq:reduced-channel-review}
\end{equation}
For hard alternatives \(a\) and \(b\), the channel acts on the off-diagonal element as
\begin{equation}
  \rho_{ab}\longmapsto \mathcal F_{ab}\rho_{ab}.
  \label{eq:coherence-map-review}
\end{equation}
The soft limit gives
\begin{equation}
  |\mathcal F_{ab}(\mu,\Lambda)|
  =
  \exp\left[-\eta_{ab}\int_\mu^\Lambda\frac{\dd\omega}{\omega}+O(1)\right]
  \sim
  \left(\frac{\mu}{\Lambda}\right)^{\eta_{ab}},
  \label{eq:dephasing-review}
\end{equation}
where \(\mu\) is an infrared floor.  The exponent vanishes when the two alternatives produce the same soft current.  It grows with the norm of their difference current.  The reduced channel measures how well the unresolved photons distinguish the hard alternatives.

I now obtain the spectral observable by retaining one property of the outcome: its total energy,
\begin{equation}
  E:X\longmapsto E(X)=\sum_{(k,\lambda)\in X}|k| .
  \label{eq:energy-map-review}
\end{equation}
The push-forward measure is
\begin{equation}
  \dd\calP_h(E)
  =
  \int \mathsf I_h[\dd X] \,\delta(E-E(X)).
  \label{eq:energy-pushforward-review}
\end{equation}
The same push-forward also defines an energy-resolved completely positive map,
\begin{equation}
  \Phi_E(\rho)\,\dd E
  =
  \int \mathcal I[\dd X](\rho)\,\delta(E-E(X))\,\dd E.
  \label{eq:energy-resolved-channel}
\end{equation}
Its Laplace transform is the tilted channel
\begin{equation}
  \Phi_\tau(\rho)
  =
  \int_0^\infty \ee^{-\tau E}\Phi_E(\rho)\,\dd E.
  \label{eq:tilted-channel}
\end{equation}
The tilted channel satisfies \(\Phi_{\tau=0}=\Phi\).  Its derivatives generate the moments of unresolved energy,
\begin{equation}
  (-1)^n\left.\frac{\partial^n}{\partial\tau^n}
  \operatorname{Tr}\Phi_\tau(\rho_h)\right|_{\tau=0}
  =\langle E^n\rangle_h .
  \label{eq:energy-moments}
\end{equation}
For a fixed diagonal hard sector, its trace gives
\begin{equation}
  \widetilde{\calP}_h(\tau)
  =
  \int_0^\infty \ee^{-\tau E}\,\dd\calP_h(E)
  =
  \exp\left[
  \int_0^\Lambda \dd N_h(\omega)(\ee^{-\tau\omega}-1)
  \right].
  \label{eq:laplace-review}
\end{equation}
The variable \(\tau\) is conjugate to total unresolved energy.  The scale \(\mu\) regulates the infrared when I impose a lower floor.  The tilted channel retains the full hard density matrix.  Its diagonal trace gives endpoint statistics; its off-diagonal matrix elements give energy-resolved coherence.

Equations \eqref{eq:dephasing-review} and \eqref{eq:laplace-review} use one instrument.  The reduced channel uses the overlap of two conditional soft environments.  The energy marginal uses the diagonal outcome distribution of one hard configuration. I now write both in one matrix formula.

\section{Energy-resolved coherence}

Let \(s_h(k,\lambda)\) be the one-photon soft amplitude density for hard configuration \(h\).  I define the sesquilinear kernel
\begin{equation}
  \dd N_{ba}(\omega)
  =
  \sum_\lambda\int \dd\Omega\,
  s_b^*(\omega,\Omega,\lambda)s_a(\omega,\Omega,\lambda)\,\dd\omega .
  \label{eq:sesquilinear-soft-kernel}
\end{equation}
The diagonal element \(\dd N_{aa}\) is the positive intensity \(\dd N_a\).  The off-diagonal element carries coherence.  In the logarithmic regime,
\begin{equation}
  \dd N_{ba}(\omega)=\kappa_{ba}\frac{\dd\omega}{\omega}
  +\dd N_{ba,{\rm reg}}(\omega).
  \label{eq:kappa-ba}
\end{equation}

I then obtain the energy-resolved coherence transform by taking the \(ba\) matrix element of the tilted channel,
\begin{equation}
  \widetilde{\calF}_{ba,\mu}(\tau)
  =
  \exp\left[
  -\frac12\int_\mu^\Lambda \dd N_{aa}
  -\frac12\int_\mu^\Lambda \dd N_{bb}
  +\int_\mu^\Lambda \ee^{-\tau\omega}\dd N_{ba}(\omega)
  \right].
  \label{eq:matrix-energy-marginal}
\end{equation}
Setting \(b=a\) gives the diagonal probability marginal,
\begin{equation}
  \widetilde{\calF}_{aa,\mu}(\tau)
  =
  \exp\left[
  \int_\mu^\Lambda \dd N_a(\omega)(\ee^{-\tau\omega}-1)
  \right]
  =\widetilde{\calP}_{a,\mu}(\tau).
  \label{eq:matrix-diagonal}
\end{equation}
Setting \(\tau=0\) gives the environmental overlap,
\begin{equation}
  \left|\widetilde{\calF}_{ba,\mu}(0)\right|
  =
  \exp\left[
  -\frac12\int_\mu^\Lambda
  \sum_\lambda\int \dd\Omega\,|s_a-s_b|^2\,\dd\omega
  \right].
  \label{eq:matrix-overlap}
\end{equation}
I obtain
\begin{equation}
  \eta_{ab}^{\rm dec}
  =
  \frac12\left(\kappa_{aa}+\kappa_{bb}-2\operatorname{Re}\kappa_{ba}\right),
  \label{eq:dephasing-kappa}
\end{equation}
while
\begin{equation}
  \eta_a^{\rm edge}=\kappa_{aa}.
  \label{eq:edge-kappa}
\end{equation}
The dephasing exponent measures the distance between two soft records.  The edge exponent measures the diagonal soft intensity of one hard configuration.

The matrix \(\kappa_{ba}\) is positive semidefinite as a Gram matrix of soft amplitudes.  Hence
\begin{equation}
  |\kappa_{ba}|^2\leq \kappa_{aa}\kappa_{bb},
  \label{eq:kappa-cauchy}
\end{equation}
and \(\eta_{ab}^{\rm dec}\geq0\).  The dephasing rate vanishes when \(s_a=s_b\) in the infrared one-particle space.

For a hard process with charged legs \(i\), the leading soft amplitude has the current form
\begin{equation}
  s_h(k,\lambda)
  =
  \sum_{i\in h}\eta_i e_i
  \frac{p_i\cdot\varepsilon_\lambda(k)}{p_i\cdot k},
  \label{eq:multileg-soft-amplitude}
\end{equation}
where \(\eta_i=+1\) for outgoing legs and \(\eta_i=-1\) for incoming legs.  The diagonal quantity \(|s_h|^2\) contains the interference terms between charged legs.  These terms contribute directly to \(\eta_h^{\rm edge}\).  The off-diagonal quantity \(|s_a-s_b|^2\) contains the corresponding difference-current interference and fixes \(\eta_{ab}^{\rm dec}\).  Gauge invariance organizes both through the conserved soft current.

The mode overlap fixes the relative signs in Eq.~\eqref{eq:matrix-energy-marginal}:
\begin{equation}
  -\frac12\dd N_{aa}-\frac12\dd N_{bb}
  \qquad\hbox{and}\qquad
  \ee^{-\tau\omega}\dd N_{ba}.
  \label{eq:real-virtual-kernel-signs}
\end{equation}
The first two terms carry the zero-event normalization.  The third term resolves an emitted soft quantum by its energy.  Both arise from one overlap before we choose a diagonal or off-diagonal projection.

At \(\tau=0\), the diagonal exponent vanishes mode by mode and enforces trace preservation.  For \(a\neq b\), the same identity leaves the negative norm \(-|s_a-s_b|^2/2\) and an environment-induced phase.  The divergent and finite pieces inherit their relative signs from this mode-level overlap.

\begin{proposition}[Soft-kernel projections]
The sesquilinear kernel \(\dd N_{ba}\) fixes both open-quantum-system observables.  Its diagonal part gives the energy marginal and the spectral edge.  Its difference-current contraction gives hard-sector dephasing.
\end{proposition}

\section{From the energy marginal to the spectral edge}

I study the inclusive charged-sector spectral measure
\begin{equation}
  \rhoinc(s)=\sum_{n=0}^{\infty}\frac{1}{n!}
  \int \dd\Phi_{e+n\gamma}(P)
  \left|\langle e(p);\gamma(k_1),\ldots,\gamma(k_n)|\psi(0)|0\rangle\right|^2
  \delta(s-P^2),
  \label{eq:rhoinc}
\end{equation}
where \(P=p+\sum_i k_i\) and \(p^2=m^2\).  This observable sums over all unresolved photon sectors at fixed invariant mass.  Every spectral statement below refers to this inclusive measure.

The field \(\psi\) serves as a charged interpolating operator.  The sum over degenerate photon states supplies the inclusive spectral weight.  I track the endpoint of this measure and avoid assigning the edge exponent to a new scale-invariant charged operator.

The instrument representation makes the energy sum explicit.  With \(\mathsf I_h[\dd X]\) normalized by
\begin{equation}
  \int \mathsf I_h[\dd X]=1,
  \label{eq:instrument-normalization}
\end{equation}
we apply
\begin{equation}
  X\longmapsto E(X)=\sum_{k\in X}\omega_k
  \label{eq:energy-map}
\end{equation}
and obtain
\begin{equation}
  \dd\calP_h(E)
  =
  \int \mathsf I_h[\dd X] \,\delta(E-E(X)).
  \label{eq:push-forward}
\end{equation}
Equation \eqref{eq:paperA-input} is the Laplace transform of this measure.

Let
\begin{equation}
  K^\mu=\sum_i k_i^\mu,\qquad P^\mu=p^\mu+K^\mu,\qquad p^2=m^2 .
  \label{eq:K-def}
\end{equation}
Then
\begin{equation}
  s-m^2=P^2-m^2=2p\cdot K+K^2 .
  \label{eq:sminusm}
\end{equation}
In the rest frame of the charge,
\begin{equation}
  s-m^2=2mE_\gamma+O(E_\gamma^2),\qquad
  E_\gamma=\sum_i\omega_i .
  \label{eq:sminusm-rest}
\end{equation}
The small-energy law of \(\calP_h\) fixes the non-analytic power in \(s-m^2\).

The map has derivative \(2m\) at the endpoint.  Its Jacobian changes the coefficient,
\begin{equation}
  \delta(s-m^2-2mE+O(E^2))
  =\frac{1}{2m}\delta\!\left(E-\frac{s-m^2}{2m}\right)+O(s-m^2),
  \label{eq:endpoint-jacobian}
\end{equation}
and preserves the exponent.

For the logarithmic term in Eq.~\eqref{eq:qed-intensity},
\begin{equation}
  \widetilde{\calP}_h(\tau)
  =
  \exp\left[
  \eta_h\int_0^\Lambda
  \frac{\dd\omega}{\omega}
  \bigl(\ee^{-\tau\omega}-1\bigr)
  \right].
  \label{eq:laplace-log}
\end{equation}
Normalization gives \(\widetilde{\calP}_h(0)=1\).  For \(\tau\Lambda\gg1\),
\begin{equation}
  \int_0^\Lambda \frac{\dd\omega}{\omega}
  \bigl(\ee^{-\tau\omega}-1\bigr)
  =
  -\log(\tau\Lambda)-\gamma_E+O((\tau\Lambda)^{-1}) .
  \label{eq:laplace-asymptotic}
\end{equation}
I find
\begin{equation}
  \widetilde{\calP}_h(\tau)
  \sim
  \ee^{-\eta_h\gamma_E}(\tau\Lambda)^{-\eta_h},
  \label{eq:laplace-power}
\end{equation}
The same cumulant has a real-time form.  Set \(\tau=\epsilon+\ii t\) with \(\epsilon>0\) and take \(\epsilon\downarrow0\).  Then
\begin{equation}
  \mathcal C_h(t)
  =
  \exp\left[
  \int_0^\Lambda\dd N_h(\omega)
  \bigl(\ee^{-\ii\omega t}-1\bigr)
  \right]
  \sim
  \ee^{-\eta_h\gamma_E}
  \ee^{-\ii\pi\eta_h/2}
  (\Lambda t)^{-\eta_h},
  \qquad t\Lambda\gg1.
  \label{eq:real-time-power}
\end{equation}
This algebraic tail is the time-domain form of the spectral edge.  Fourier transformation converts \(t^{-\eta_h}\) into the positive-energy law \(E^{\eta_h-1}\).  The Laplace and real-time forms encode one endpoint singularity.

The inverse Laplace transform gives
\begin{equation}
  \calP_h(E)
  \sim
  \frac{\ee^{-\eta_h\gamma_E}}{\Gamma(\eta_h)}
  \frac{E^{\eta_h-1}}{\Lambda^{\eta_h}},
  \qquad E\downarrow0 .
  \label{eq:energy-edge}
\end{equation}

I collect hard and angular factors into \(H(E)\):
\begin{equation}
  \rhoinc(s)
  =
  \int_0^\infty \dd E\,H(E)\calP_h(E)
  \delta\bigl(s-m^2-2mE+O(E^2)\bigr).
  \label{eq:rho-from-P}
\end{equation}
For the one-electron spectral measure,
\begin{equation}
  H(E)=H(0)+O(E),
  \qquad H(0)\neq0 .
  \label{eq:H-smooth}
\end{equation}
I obtain
\begin{equation}
  \rhoinc(s)
  \sim
  C\,\theta(s-m^2)(s-m^2)^{-1+\eta_h}.
  \label{eq:rho-edge}
\end{equation}

\begin{theorem}[Soft-instrument edge]
Let the diagonal energy marginal have infrared intensity
\(\eta_h\dd\omega/\omega\) plus an integrable measure, and let \(H(0)\neq0\).  Then the inclusive charged spectral measure obeys Eq.~\eqref{eq:rho-edge}.  Its two-point function has a branch point at \(p^2=m^2\) for non-integer \(\eta_h\).
\end{theorem}

Fixed-order perturbation theory expands the power as
\begin{equation}
  (s-m^2)^{-1+\eta_h}
  =
  (s-m^2)^{-1}
  \left[1+\eta_h\log(s-m^2)+O(\eta_h^2)\right].
  \label{eq:fixed-order-expansion}
\end{equation}
The expansion is understood as an endpoint distribution.  As \(\eta_h\to0^+\), the normalized edge kernel returns the delta function; Appendix~A gives the test-function limit.  For finite \(\eta_h\), the resummed cumulant turns the logarithmic series into a branch point.  The endpoint contribution to the two-point function reads
\begin{equation}
  G_{\rm edge}(p^2)
  \simeq
  \int_{m^2}^{m^2+\epsilon}
  \frac{\dd s\,C(s-m^2)^{-1+\eta_h}}{p^2-s+\ii0}.
  \label{eq:edge-propagator}
\end{equation}
For non-integer \(\eta_h\), \(p^2=m^2\) is a branch point and the isolated pole weight vanishes.

\section{Unresolved-photon sectors}

I separate the Laplace variable \(\tau\) from the infrared floor \(\mu\).  With \(\mu>0\),
\begin{equation}
  \widetilde{\calP}_{h,\mu}(\tau)
  =
  \exp\left[
  \int_\mu^\Lambda \dd N_h(\omega)
  \bigl(\ee^{-\tau\omega}-1\bigr)
  \right].
  \label{eq:laplace-mu}
\end{equation}
Expanding the exponential gives
\begin{align}
  \widetilde{\calP}_{h,\mu}(\tau)
  &=
  \exp\left[-\int_\mu^\Lambda \dd N_h(\omega)\right]
  \sum_{n=0}^{\infty}\frac{1}{n!}
  \prod_{i=1}^n
  \left[\int_\mu^\Lambda \dd N_h(\omega_i)\right]
  \exp\left[-\tau\sum_{i=1}^n\omega_i\right].
  \label{eq:real-virtual-series}
\end{align}
The prefactor is the zero-photon probability.  The \(n\)-th term contains \(n\) real unresolved photons.  Their total energy enters through the last exponential.  At \(\tau=0\),
\begin{equation}
  \sum_{n=0}^\infty P_n(\mu)=1,
  \qquad
  P_0(\mu)=\exp\left[-\int_\mu^\Lambda\dd N_h\right].
  \label{eq:sector-normalization}
\end{equation}
The logarithmic bath sends \(P_0\) to zero and transfers its weight to sectors with real unresolved photons.

The energy density in the \(n\)-photon sector is
\begin{equation}
  \calP_{n,\mu}(E)
  =
  \ee^{-N_\mu}\frac{1}{n!}
  \int_\mu^\Lambda\prod_{i=1}^n\dd N_h(\omega_i)
  \delta\!\left(E-\sum_{i=1}^n\omega_i\right),
  \qquad
  N_\mu=\int_\mu^\Lambda\dd N_h.
  \label{eq:Pn-energy-density}
\end{equation}
The full distribution is \(\calP_{h,\mu}(E)=\sum_{n=0}^\infty\calP_{n,\mu}(E)\).  Convolution in energy becomes multiplication in Laplace space, which explains the exponential form of Eq.~\eqref{eq:laplace-mu}.

A counting variable \(z\) keeps photon number and total energy in one generator:
\begin{equation}
  \mathcal Z_{h,\mu}(z,\tau)
  =
  \exp\left[
  \int_\mu^\Lambda\dd N_h(\omega)
  \bigl(z\ee^{-\tau\omega}-1\bigr)
  \right].
  \label{eq:number-energy-generator}
\end{equation}
At \(z=1\) it reduces to the energy marginal.  At \(\tau=0\) it generates the photon-number distribution.  Derivatives with respect to \(z\) give factorial moments, while derivatives with respect to \(\tau\) give energy moments.

The logarithmic bath separates multiplicity from energy.  Its mean photon number is
\begin{equation}
  \langle n\rangle_\mu
  =
  \int_\mu^\Lambda\eta_h\frac{\dd\omega}{\omega}
  =
  \eta_h\log\frac{\Lambda}{\mu},
  \label{eq:mean-soft-number}
\end{equation}
which diverges as \(\mu\to0\).  Its mean unresolved energy remains finite,
\begin{equation}
  \langle E\rangle
  =
  \int_0^\Lambda \omega\,\eta_h\frac{\dd\omega}{\omega}
  =
  \eta_h\Lambda,
  \label{eq:mean-soft-energy}
\end{equation}
within the soft approximation and cutoff convention used here.  The infraparticle edge combines infinite soft multiplicity with finite total unresolved energy.

The inclusive transform has a finite \(\mu\to0\) limit because
\begin{equation}
  \frac{\ee^{-\tau\omega}-1}{\omega}
  =
  -\tau+O(\omega),
  \qquad \omega\downarrow0 .
  \label{eq:tau-small-energy}
\end{equation}
The zero-photon probability vanishes, while the sum over real sectors preserves unit normalization.  Large \(\tau\) probes small total energy, while \(\mu\) sets the infrared floor.

At fixed photon number, the energy constraint is
\begin{equation}
  \delta\left(E-\sum_{i=1}^n\omega_i\right).
  \label{eq:n-sector-energy}
\end{equation}
Each fixed-\(n\) sector contributes to the endpoint.  The noninteger power follows from their linked all-orders sum.

\section{Hard threshold factor}

Equation \eqref{eq:rho-from-P} factorizes the endpoint into the soft marginal \(\calP_h(E)\) and the hard factor \(H(E)\).  The one-electron problem has the regular form Eq.~\eqref{eq:H-smooth}.  To parameterize another threshold observable, write
\begin{equation}
  H(E)\sim E^{-\chi}H_0(E),
  \qquad H_0(0)\neq0.
  \label{eq:H-chi}
\end{equation}
For the one-electron measure,
\begin{equation}
  \chi_{\rm e}=0.
  \label{eq:chi-min}
\end{equation}
A regular threshold selection rule can also give \(H(E)\sim E^rH_0(E)\) with \(r>0\); this shifts the ordinary threshold power by \(r\) while leaving the soft contribution \(\eta_h\) intact.  A singular hard factor shifts the total power:
\begin{equation}
  \rho(s)
  \sim
  C_\chi\,\theta(s-m^2)(s-m^2)^{-1+\eta_h-\chi}.
  \label{eq:rho-edge-chi}
\end{equation}
The two coefficients have separate origins.  The soft instrument fixes \(\eta_h\).  The measured hard threshold fixes \(\chi\).

This separation also fixes the order of operations.  I first integrate the unresolved photon instrument and obtain \(\calP_h(E)\).  I then combine it with the hard threshold factor in Eq.~\eqref{eq:rho-from-P}.  Soft-photon resummation controls \(\eta_h\); threshold dynamics of the measured hard operator controls \(\chi\).  The one-electron problem isolates the soft contribution because \(\chi_{\rm e}=0\).

\begin{proposition}[Separation of exponents]
For \(\calP_h(E)\sim E^{-1+\eta_h}\) and \(H(E)\sim E^{-\chi}H_0(E)\), the endpoint exponent is \(-1+\eta_h-\chi\).  The one-electron spectral measure has \(\chi=0\).
\end{proposition}

\section{Comparison with unparticle spectra}

I write the QED edge in the form
\begin{equation}
  \rhoinc(s)\propto
  \theta(s-m^2)(s-m^2)^{\dEff-2},
  \qquad
  \dEff=1+\eta_h .
  \label{eq:deff}
\end{equation}
Here \(\dEff\) labels the endpoint power.  The electron retains its mass \(m\), and the scale-free factor comes from the unresolved photon measure.  By comparison, \(\dU\) in Eq.~\eqref{eq:georgi-rho} is an operator dimension of a scale-invariant sector.  Three properties identify the QED exponent as environmental: it depends on the charged hard configuration, it adds over independent unresolved sectors, and it controls the flow of the finite-resolution atomic weight.

The support also retains the mass scale: \(\rhoinc(s)=0\) below \(m^2\).  The soft bath changes the analytic type at that threshold but leaves the threshold location fixed.  This gives a massive edge with a scale-free local profile.

The infrared floor separates the two constructions.  In QED, a floor \(\mu\) leaves an atomic weight
\begin{equation}
  Z(\mu)
  =
  \exp\left[
  -\int_\mu^\Lambda
  \eta_h\frac{\dd\omega}{\omega}
  \right]
  =
  \left(\frac{\mu}{\Lambda}\right)^{\eta_h}.
  \label{eq:residue-flow}
\end{equation}
Removing the floor gives
\begin{equation}
  \lim_{\mu\to0}Z(\mu)=0 .
  \label{eq:Zzero}
\end{equation}
The edge is the zero-resolution limit of a massive threshold whose atom loses all weight.

The cumulative edge weight is
\begin{equation}
  W(\Delta E)
  =
  \int_0^{\Delta E}\calP_h(E)\,\dd E
  \sim
  \frac{\ee^{-\eta_h\gamma_E}}{\Gamma(1+\eta_h)}
  \left(\frac{\Delta E}{\Lambda}\right)^{\eta_h}.
  \label{eq:cumulative-weight}
\end{equation}
Hence
\begin{equation}
  \frac{W(a\Delta E)}{W(\Delta E)}\to a^{\eta_h},
  \qquad \Delta E\downarrow0 .
  \label{eq:ratio-law}
\end{equation}
This ratio measures the soft response coefficient.

Independent unresolved sectors multiply their Laplace transforms:
\begin{equation}
  \widetilde{\calP}_{h,{\rm tot}}(\tau)
  =
  \prod_a \widetilde{\calP}_{h,a}(\tau)
  \sim
  C\,\tau^{-\sum_a\eta_{h,a}} .
  \label{eq:additive-laplace}
\end{equation}
This gives
\begin{equation}
  \dEff-1=\sum_a\eta_{h,a}.
  \label{eq:index-additivity}
\end{equation}
The endpoint index adds because independent environmental cumulants add.

The rule follows at the level of probability laws: total unresolved energy is the sum of independent sector energies, so their distributions convolve and their Laplace transforms multiply.  The exponent records this convolution law.

Define
\begin{equation}
  W(E)=\int_0^E \calP_h(E')\,\dd E' .
  \label{eq:WE-def}
\end{equation}
For QED,
\begin{equation}
  W(E)\sim
  \frac{\ee^{-\eta_h\gamma_E}}{\Gamma(1+\eta_h)}
  \left(\frac{E}{\Lambda}\right)^{\eta_h},
  \qquad E\downarrow0 .
  \label{eq:WE-asymptotic}
\end{equation}
We can extract
\begin{equation}
  \eta_h=
  \lim_{E\downarrow0}
  \frac{\log W(aE)-\log W(E)}{\log a},
  \qquad a>0 .
  \label{eq:eta-resolution}
\end{equation}
More generally,
\begin{equation}
  W(E)\sim B E^\eta L(E),
  \qquad E\downarrow0,
  \label{eq:regularly-varying-W}
\end{equation}
with \(L\) slowly varying, gives
\begin{equation}
  \widetilde{\calP}(\tau)\sim
  B\Gamma(1+\eta)\tau^{-\eta}L(1/\tau),
  \qquad \tau\to\infty .
  \label{eq:tauberian}
\end{equation}
Within the compound soft-instrument form Eq.~\eqref{eq:general-generator}, the measured index equals the coefficient of \(\dd\omega/\omega\) when the intensity contains no stronger infrared term.  The power law determines this coefficient inside that class.  Microscopic reconstruction requires the full kernel \(\dd N_{ba}\), including its angular, polarization, and hard-state dependence.

\begin{proposition}[Resolution extraction]
For a compound soft marginal with logarithmic leading intensity, the limit Eq.~\eqref{eq:eta-resolution} returns \(\eta_h\).
\end{proposition}

\section{Infrared classes of the energy marginal}

Consider
\begin{equation}
  \widetilde{\calP}_h(\tau)
  =
  \exp\left[
  \int_0^\Lambda
  \dd N_h(\omega)
  \bigl(\ee^{-\tau\omega}-1\bigr)
  \right].
  \label{eq:general-generator}
\end{equation}
The infrared behavior of \(\dd N_h\) gives three endpoint classes.  The classification follows from the total soft intensity near \(\omega=0\), so it applies directly to the energy marginal of the open-system instrument.

For an integrable intensity,
\begin{equation}
  \calP_h(E)
  =
  \ee^{-N_{\rm tot}}\delta(E)+\calP_{h,{\rm cont}}(E),
  \qquad
  N_{\rm tot}=\int_0^\Lambda\dd N_h .
  \label{eq:atom-finite-bath}
\end{equation}
The atom retains weight \(\ee^{-N_{\rm tot}}\).

This weight is the probability of no unresolved quantum.  A finite total intensity leaves that probability nonzero even after the infrared floor is removed.

For \(\dd N_h=\eta_h\dd\omega/\omega\), the atom loses all weight and the continuum has a power edge.  QED lies in this class.

The total intensity now grows as \(\eta_h\log(\Lambda/\mu)\), so the no-photon probability scales as \((\mu/\Lambda)^{\eta_h}\).  At the same time, the subtracted Laplace exponent grows as \(-\eta_h\log(\tau\Lambda)\), which produces the power edge.  QED lies in this class.

For a stronger singularity,
\begin{equation}
  \dd N_\alpha(\omega)
  =
  \kappa\frac{\dd\omega}{\omega}
  \left(\frac{\Lambda}{\omega}\right)^\alpha,
  \qquad 0<\alpha<1,
  \label{eq:more-singular}
\end{equation}
we obtain
\begin{equation}
  \log\widetilde{\calP}_\alpha(\tau)
  \sim
  -A_\alpha(\tau\Lambda)^\alpha,
  \qquad A_\alpha>0 .
  \label{eq:stretched}
\end{equation}
The atom again vanishes, while the endpoint ceases to be a simple power law.

\begin{theorem}[Endpoint classes]
An integrable soft intensity retains an atom.  A logarithmic intensity produces a power-law edge.  A stronger nonintegrable intensity removes the atom and produces a non-power endpoint.
\end{theorem}

The subtraction in Eq.~\eqref{eq:general-generator} preserves total probability in every class.  For the logarithmic class, the zero-energy probability vanishes and the continuum carries the full weight.  Inclusive infrared cancellation and pole loss occur together.

Now add an integrable measure:
\begin{equation}
  \dd N_h(\omega)
  =
  \eta_h\frac{\dd\omega}{\omega}
  +
  \dd N_{h,{\rm reg}}(\omega),
  \qquad
  \int_0^\Lambda\dd N_{h,{\rm reg}}<\infty .
  \label{eq:log-plus-reg}
\end{equation}
Then
\begin{equation}
  \widetilde{\calP}_h(\tau)
  =
  \widetilde{\calP}_{h,\log}(\tau)
  \exp\left[
  \int_0^\Lambda
  \dd N_{h,{\rm reg}}(\omega)
  \bigl(\ee^{-\tau\omega}-1\bigr)
  \right].
  \label{eq:regular-factor}
\end{equation}
The second factor approaches a finite constant and changes the normalization and sub-leading terms.  The power remains \(\eta_h\).

\begin{proposition}[Stability]
Infrared-integrable additions to the intensity preserve the leading power \((s-m^2)^{-1+\eta_h}\).
\end{proposition}

\section{Finite resolution and residue flow}

Impose an infrared floor \(\mu\):
\begin{equation}
  \dd N_{h,\mu}(\omega)
  =
  \eta_h\frac{\dd\omega}{\omega}\,
  \theta(\omega-\mu)\theta(\Lambda-\omega).
  \label{eq:finite-floor-intensity}
\end{equation}
The zero-photon weight is
\begin{equation}
  Z(\mu)
  =
  \exp\left[
  -\int_\mu^\Lambda
  \eta_h\frac{\dd\omega}{\omega}
  \right]
  =
  \left(\frac{\mu}{\Lambda}\right)^{\eta_h}.
  \label{eq:finite-floor-residue}
\end{equation}
It obeys
\begin{equation}
  \lim_{\mu\to0}Z(\mu)=0
  \label{eq:mu-limit}
\end{equation}
and
\begin{equation}
  \mu\frac{\dd Z(\mu)}{\dd\mu}=\eta_h Z(\mu).
  \label{eq:Z-rg}
\end{equation}

\begin{proposition}[Residue flow]
The endpoint exponent \(\eta_h\) generates the resolution flow of the atomic weight through Eq.~\eqref{eq:Z-rg}.
\end{proposition}

A detector supplies a nonzero effective \(\mu\) through finite time, finite volume, energy resolution, thresholds, or screening.  A finite observation time \(T\) gives \(\mu\sim T^{-1}\); a finite box of size \(L\) gives \(\mu\sim L^{-1}\); an energy threshold gives \(\mu\sim\Delta E\).  At that scale the charged excitation has a finite operational residue.  Sending \(\mu\) to zero moves that weight into the continuum without changing the mass threshold.

At that scale the charged excitation has a finite operational residue.  Lowering \(\mu\) refines the environmental coarse graining and opens more unresolved photon sectors.  The hard charge and its threshold remain fixed while the atomic weight flows into those sectors.

\section{Infrared cancellation and decoherence}

Bloch--Nordsieck, Yennie--Frautschi--Suura, Kinoshita, and Lee--Nauenberg cancellation make inclusive probabilities finite \cite{BlochNordsieck1937,YennieFrautschiSuura1961,Kinoshita1962,LeeNauenberg1964,Weinberg1965}.  The normalized energy marginal implements that cancellation through
\(\ee^{-\tau\omega}-1\).  Its total weight remains one, while its exact zero-energy weight vanishes in the logarithmic limit.  The cancellation removes the regulator from the inclusive distribution.  In the logarithmic class the delta coefficient vanishes and the continuum carries the full weight.

The reduced channel and the spectral edge use the same kernel.  The channel gives
\begin{equation}
  \exp\left[
  -\eta_{ab}
  \int_{\omega_{\IR}}^\Lambda
  \frac{\dd\omega}{\omega}
  \right]
  =
  \left(\frac{\omega_{\IR}}{\Lambda}\right)^{\eta_{ab}}
  \label{eq:decoherence-factor}
\end{equation}
for an off-diagonal hard coherence.  The diagonal energy marginal gives Eq.~\eqref{eq:paperA-input} and the spectral edge.  The first observable measures the distinguishability of two soft records.  The second measures the energy distribution of one soft record.

Real and virtual terms retain the same current kernel in both observables.  The zero-event pieces carry negative half-norms; the resolved-event piece carries the positive cross kernel.  Angular integration and hard kinematics then generate the correlated logarithmic and finite terms found in perturbation theory.

This relation places the infraparticle edge inside the open-system framework.  The environment dephases superpositions of hard charged alternatives and dissolves the diagonal spectral atom.  The coefficients differ according to Eq.~\eqref{eq:dephasing-kappa} and Eq.~\eqref{eq:edge-kappa}; the underlying soft kernel is common.

\section{Conclusion}

In this paper, built upon open-quantum-system formulation of soft-QED~\cite{PaperA,PaperB,PaperC}, I derived the infraparticle edge behavior from the energy marginal of the soft-QED quantum instrument.  The logarithmic intensity
\begin{equation}
  \dd N_h(\omega)=\eta_h\frac{\dd\omega}{\omega}+\dd N_{h,{\rm reg}}(\omega)
\end{equation}
gives
\begin{equation}
  \rhoinc(s)\sim
  C\,\theta(s-m^2)(s-m^2)^{-1+\eta_h}.
\end{equation}
The same instrument also fixes hard-sector dephasing.  Its diagonal contraction gives \(\eta_h^{\rm edge}=\kappa_{hh}\); its difference-current contraction gives \(\eta_{ab}^{\rm dec}\).  This matrix structure joins reduced dynamics and endpoint spectroscopy without identifying their numerical exponents.

The finite-resolution atom obeys
\[
  Z(\mu)=\left(\frac{\mu}{\Lambda}\right)^{\eta_h},
\]
and independent unresolved sectors add their logarithmic coefficients.  Integrable intensities retain an atom, logarithmic intensities produce power edges, and stronger intensities produce non-power endpoints.  A regular hard factor leaves the one-electron exponent unchanged.  A singular hard threshold contributes a separate shift \(\chi\).  We obtain a direct separation between the open-system bath response and the measured hard threshold.

The resulting continuum has the local power law of a gapped unparticle spectrum.  In QED, the power comes from unresolved photons around a massive charge.  The electron remains massive, and the scale-free structure resides in the environment.

These results are unique characteristics of soft-QED. I foresee immediate applications for a diverse multi-scale phenomena. 

\section*{Acknowledgments}
I acknowledge numerous feedbacks after my talk on this work at workshops at the Institute for Nuclear Theory (December 2025, USA) and at the Mainz Institute for Theoretical Physics (April 2026, Germany). Part of this work was performed while I was visiting the Institute for Theoretical Physics at Cologne University (Germany), the Institute for Pure and Applied Mathematics (IPAM, USA), the Simons Institute for the Theory of Computing (SIfTC, USA), and the Fields Institute for Research in Mathematical Sciences (FIRMS, Canada). This work was supported in part by the DFG Cluster of Excellence PRISMA+ (Project ID 39083469), by the U.S. National Science Foundation through IPAM and SIfTC, by the Simons Foundation through SIfTC, and by the National Research Foundation of Korea (NRF) (RS-2021-NR060112) and Kwangwoon University. 

\appendix

\section{Asymptotic Laplace transform}

Set
\begin{equation}
  I(\tau)=
  \int_0^\Lambda\frac{\dd\omega}{\omega}
  \bigl(\ee^{-\tau\omega}-1\bigr).
\end{equation}
Differentiation gives
\begin{equation}
  I'(\tau)
  =
  -\int_0^\Lambda \ee^{-\tau\omega}\dd\omega
  =
  -\frac{1-\ee^{-\tau\Lambda}}{\tau}.
\end{equation}
Integration yields
\begin{equation}
  I(\tau)
  =
  -\log(\tau\Lambda)-\gamma_E+O((\tau\Lambda)^{-1})
\end{equation}
for \(\tau\Lambda\gg1\).  Hence
\begin{equation}
  \widetilde{\calP}_h(\tau)
  \sim
  \ee^{-\eta_h\gamma_E}(\tau\Lambda)^{-\eta_h}.
\end{equation}
We invert it with
\begin{equation}
  \int_0^\infty \ee^{-\tau E}E^{\eta_h-1}\dd E
  =
  \Gamma(\eta_h)\tau^{-\eta_h}.
\end{equation}

For the stronger measure Eq.~\eqref{eq:more-singular}, \(x=\tau\omega\) gives
\begin{equation}
  \log\widetilde{\calP}_\alpha(\tau)
  \sim
  \kappa(\tau\Lambda)^\alpha
  \int_0^\infty
  \frac{\dd x}{x^{1+\alpha}}
  (\ee^{-x}-1)
  =
  -A_\alpha(\tau\Lambda)^\alpha .
\end{equation}
The logarithmic measure is the unique boundary between a finite atom and a non-power endpoint within this family.

The limit \(\eta_h\to0^+\) restores the atom.  For a smooth test function \(f\),
\begin{equation}
  \lim_{\eta_h\to0^+}
  \frac{1}{\Gamma(\eta_h)}
  \int_0^\infty E^{\eta_h-1}f(E)\dd E
  =
  f(0).
\end{equation}
Thus
\(\Gamma(\eta_h)^{-1}E^{\eta_h-1}\theta(E)\to\delta(E)\).

\section{Real-time asymptotics}

For \(t>0\), define
\begin{equation}
  I_t=
  \int_0^\Lambda\frac{\dd\omega}{\omega}
  \bigl(\ee^{-\ii\omega t}-1\bigr).
\end{equation}
Analytic continuation of the Laplace result gives
\begin{equation}
  I_t
  =
  -\log(\Lambda t)-\gamma_E-\frac{\ii\pi}{2}
  +O((\Lambda t)^{-1}).
\end{equation}
Hence \(\mathcal C_h(t)=\exp[\eta_h I_t]\) obeys Eq.~\eqref{eq:real-time-power}.  The phase depends on the time-ordering convention, while the decay power is fixed by \(\eta_h\).  The Fourier integral
\begin{equation}
  \int_0^\infty \dd t\,\ee^{\ii Et}t^{-\eta_h}
  =
  \ee^{\ii\pi(1-\eta_h)/2}
  \Gamma(1-\eta_h)E^{\eta_h-1}
\end{equation}
for \(0<\eta_h<1\) displays the endpoint power obtained from the Laplace inversion.

\section{Regular variation}

Let
\[
  W(E)\sim B E^\eta L(E)
\]
with \(L\) slowly varying at the origin.  Then
\[
  \widetilde{\calP}(\tau)
  =
  \int_0^\infty \ee^{-\tau E}\,\dd W(E)
  \sim
  B\Gamma(1+\eta)\tau^{-\eta}L(1/\tau).
\]
The logarithmic derivative gives
\[
  -\frac{\dd\log\widetilde{\calP}}{\dd\log\tau}\to\eta,
\]
and the cumulative ratio gives
\[
  \frac{W(aE)}{W(E)}\to a^\eta .
\]
For the compound generator Eq.~\eqref{eq:general-generator}, this index equals the coefficient of the logarithmic intensity when stronger infrared terms are absent.

\section{Threshold kinematics}

With \(P=p+K\), \(p^2=m^2\), and future-directed soft momentum \(K\),
\[
  P^2-m^2=2p\cdot K+K^2 .
\]
In the rest frame of \(p\),
\[
  P^2-m^2=2mE_\gamma+O(E_\gamma^2).
\]
The derivative at \(E_\gamma=0\) is nonzero.  The map changes the normalization and preserves the endpoint exponent.

\section{Additivity}

For independent unresolved sectors,
\begin{equation}
  \widetilde{\calP}_{\rm tot}(\tau)
  =
  \prod_a \widetilde{\calP}_a(\tau).
\end{equation}
If
\begin{equation}
  \widetilde{\calP}_a(\tau)
  \sim C_a\tau^{-\eta_a},
\end{equation}
then
\begin{equation}
  \widetilde{\calP}_{\rm tot}(\tau)
  \sim
  \left(\prod_a C_a\right)
  \tau^{-\sum_a\eta_a},
\end{equation}
and
\begin{equation}
  \eta_{\rm tot}=\sum_a\eta_a .
\end{equation}

\section{Infrared-integrable perturbations}

Let
\(\dd N=\eta\dd\omega/\omega+\dd N_{\rm reg}\), with
\(\int_0^\Lambda\dd N_{\rm reg}<\infty\).  Then
\begin{equation}
  \log\widetilde{\calP}(\tau)
  =
  \eta I(\tau)
  +
  \int_0^\Lambda\dd N_{\rm reg}(\omega)(\ee^{-\tau\omega}-1).
\end{equation}
For a bounded density at the origin, the second term equals
\(-N_{\rm reg}+O(\tau^{-1})\).  We obtain
\begin{equation}
  \widetilde{\calP}(\tau)
  \sim
  \ee^{-N_{\rm reg}}\ee^{-\eta\gamma_E}(\tau\Lambda)^{-\eta}
  \left[1+O(\tau^{-1})\right].
\end{equation}
The integrable measure changes the coefficient and subleading powers.

An integrable measure with weaker regularity can produce a slowly varying subleading factor.  Its contribution remains bounded at large \(\tau\), so it leaves the coefficient of \(\log\tau\) unchanged.

\section{Photon-number expansion}

Define
\begin{equation}
  N_\mu=\int_\mu^\Lambda \dd N_h(\omega).
\end{equation}
The zero-photon probability is \(\ee^{-N_\mu}\).  The \(n\)-photon measure is
\begin{equation}
  \dd P_n
  =
  \ee^{-N_\mu}\frac{1}{n!}
  \prod_{i=1}^n \dd N_h(\omega_i).
\end{equation}
Weighting by total energy gives
\begin{equation}
  \widetilde{\calP}_{h,\mu}(\tau)
  =
  \sum_{n=0}^{\infty}
  \ee^{-N_\mu}\frac{1}{n!}
  \prod_{i=1}^n
  \left[\int_\mu^\Lambda \dd N_h(\omega_i)\ee^{-\tau\omega_i}\right].
\end{equation}
The sum gives
\begin{equation}
  \widetilde{\calP}_{h,\mu}(\tau)
  =
  \exp\left[
  \int_\mu^\Lambda \dd N_h(\omega)(\ee^{-\tau\omega}-1)
  \right].
\end{equation}
For \(\dd N_h=\eta_h\dd\omega/\omega\), the zero-photon factor scales as
\((\mu/\Lambda)^{\eta_h}\), while the inclusive transform stays finite at fixed \(\tau>0\).

\section{Coherence kernel}

For one soft mode, the conditional amplitudes \(s_a\) and \(s_b\) give
\begin{equation}
  \exp\left[
  -\frac12 |s_a|^2
  -\frac12 |s_b|^2
  +\ee^{-\tau\omega}s_b^*s_a
  \right].
  \label{eq:single-mode-overlap}
\end{equation}
Multiplying over modes gives Eq.~\eqref{eq:matrix-energy-marginal}.  For \(a=b\), the exponent becomes
\(|s_a|^2(\ee^{-\tau\omega}-1)\).  For \(\tau=0\),
\begin{equation}
  -\frac12 |s_a|^2-\frac12 |s_b|^2+s_b^*s_a
  =
  -\frac12|s_a-s_b|^2
  +\ii\,\operatorname{Im}(s_b^*s_a).
  \label{eq:overlap-identity}
\end{equation}
The real part gives dephasing.  The imaginary part gives the environment-induced phase.

\section{Multi-leg soft coefficient}

The leading soft current of a hard configuration is
\begin{equation}
  j_h^\mu(k)
  =
  \sum_{i\in h}\eta_i e_i\frac{p_i^\mu}{p_i\cdot k}.
\end{equation}
The one-photon amplitude is \(s_h(k,\lambda)=\varepsilon_{\lambda\mu}(k)j_h^\mu(k)\).  The diagonal intensity is
\begin{equation}
  \dd N_h(\omega)
  =
  \sum_\lambda
  \frac{\omega^2\dd\Omega\,\dd\omega}{(2\pi)^3 2\omega}
  \left|\varepsilon_\lambda(k)\cdot j_h(k)\right|^2.
\end{equation}
Since \(j_h(k)\propto\omega^{-1}\), the radial measure gives
\begin{equation}
  \dd N_h(\omega)=\eta_h\frac{\dd\omega}{\omega},
\end{equation}
with
\begin{equation}
  \eta_h
  =
  \frac{1}{2(2\pi)^3}
  \sum_\lambda\int\dd\Omega\,
  \left|\omega\,\varepsilon_\lambda(k)\cdot j_h(k)\right|^2.
\end{equation}
The square contains diagonal leg terms and interference terms.  For two hard configurations, replacing \(|s_h|^2\) by \(s_b^*s_a\) gives \(\kappa_{ba}\); replacing it by \(|s_a-s_b|^2/2\) gives the dephasing coefficient.

Current conservation, \(k_\mu j_h^\mu(k)=0\), removes gauge-dependent polarization components.  The conserved current controls the diagonal intensity, the off-diagonal coherence kernel, and the real--virtual cancellation in the energy marginal.

\end{document}